\documentstyle[namedreferences,epsf]{spackap}

\begin{opening}
\title{Element Settling in the Solar Interior}

\author{Sylvie \surname{Vauclair}}
\institute{Laboratoire d'Astrophysique\ Observatoire Midi-Pyr\'en\'ees
\ Toulouse, France}

\end{opening}

\runningauthor{Sylvie Vauclair}
\runningtitle{Element Settling in the Sun}

\begin{document}


\begin{abstract}

Element settling inside the Sun now becomes detectable from the
comparison of the observed oscillation modes with the results of the
theoretical models. This settling is due, not only to gravitation, but
also to thermal diffusion and radiative acceleration (although this last
effect is small compared to the two others ). It leads to abundance
variations of helium and heavy elements of $\cong~10\%$ below the
convective zone. Although not observable from spectroscopy, such
variations lead to non-negligible modifications of the solar internal
structure and evolution. Helioseismology is a powerful tool to detect
such effects, and its positive results represent a great  success for
the theory of stellar evolution.
Meanwhile, evidences are obtained that the element settling is slightly
smoothed down, probably due to mild macroscopic motions below the
convective zone. Additional observations of the abundances of both
$^7$Li and $^3$He lead to specific constraints on these particular
motions.

\end{abstract}


\section{Introduction}

The importance of element settling inside the stars during their
evolution is now widely recognized as a ``standard process" (see, for
example, Vauclair 1998). As soon as condensed from interstellar clouds,
the self-gravitating spheres built density, pressure
and temperature gradients which force the various
chemical species present in the stellar gas to move with respect to one
another. This process, first introduced by Michaud 1970
to account for the chemically peculiar stars, 
is believed to be the reason for the large
abundance variations 
observed in main-sequence
type stars,  horizontal branch stars and white dwarfs (Vauclair and
Vauclair, 1982).

Inside the convective regions, the rapid macroscopic motions mix the gas
components and force their homogenisation. The chemical
composition observed in the external regions of cool stars is thus
affected by the settling which occurs below the outer convective zones.
As the settling time scales vary in first approximation like the inverse
of the density, the expected variations are smaller for cooler stars,
which have deeper convection zones. While some elements can see their
abundances vary by several orders of magnitude in the hottest Ap stars,
the maximum expected variations in the Sun are not larger than 
$\cong~10\%$.

Such variations cannot be observed in the solar atmosphere by
spectroscopy. 
In the present days however, due to helioseismology, we know the
internal
structure of the Sun with a high degree of precision. 
Evidences for
the
occurence of element settling are found. 
Abundance variations of the order of a few percent now become indirectly
detectable, by comparisons of the theoretical computations with the
results of the inversion of pulsating modes. We have entered a new area
in this respect.

\section{Theory of element settling}

\subsection{The Diffusion Equation}

What we use to call ``microscopic" diffusion of the chemical elements
in stars represents a competition between two kinds of processes.
First the individual atoms want to move under the influence of 
the local gravity (or pressure gradient), thermal gradient, radiative
acceleration and concentration gradient. Second their motion is slowed
down due to collisions with the other ions as they
share the acquired momentum in a random way. This competition leads to
selective element settling inside the stars.

The computations of this settling process are based on the Boltzmann
equation for dilute collision-dominated plasmas. At equilibrium the
solution of the equation is the maxwellian distribution function.
We consider here situations where the distribution is not maxwellian,
but where the deviations from the maxwellian distribution are very
small.

Two different methods have been used to solve the Boltzmann equation in
the framework of this approximation. The first method relies on the
Chapman-Enskog procedure (described in Chapman and Cowling, 1970), using
convergent series of the distribution function. This
procedure is applied to binary mixtures, leading to expressions with
successive approximations for the binary
diffusion coefficients. For the diffusion of charged particles in a
plasma, a ternary mixture approximation is introduced, including the
electrons. This method was widely used in the first computations of
diffusion processes in stars (see Vauclair and Vauclair, 1982). More
recently, similar methods have still been used by many authors, for
example Bahcall and Loeb (1990), Proffitt and Michaud (1991), Michaud
and
Vauclair (1991), Bahcall and Pinsonneault (1992), 
Charbonnel, Vauclair, Zahn (1992), Richard et al (1996) (hereafter
RVCD).
The second method is that of Burgers (1969), in which separate flow and
heat equations for each component of a multi-component mixture are
solved simultaneously. Descriptions of this method may be found for
example in Cox,
Guzik and Kidman (1989), Proffitt and VandenBerg (1991), Richer and
Michaud (1993), Thoul, Bahcall and Loeb (1994) .

In the formalism of RVCD, which has been used for the results given
below, the local abundances of the elements are given in terms of their
concentrations, solutions of equations of the following type :

$$
  \frac{\partial c_i}{\partial t} = D'_{1i} \frac{\partial ^2
c_i}{\partial
  m_r ^2} + \left( \frac{\partial D'_{1i}}{\partial m_r} - V'_{1i}
\right)
  \frac{\partial c_i}{\partial m_r} - \left( \frac{\partial
  V'_{1i}}{\partial m_r} + \lambda_i \right) c_i
$$

where $c_i$, the concentration of element i, is given in terms of
the stellar mass fraction $m_r$, 
$\lambda_i$ is the nuclear
reaction rate, and $D'_{1i}$ is given by:

$$
D'_{1i} = \left( 4 \pi \rho r^2 \right) ^2 \left( D_T+D_{1i} \right)
$$

in which $D_T$ represents the effective macroscopic diffusion 
coefficient and $D_{1i}$ the microscopic diffusion coefficient of
element i relative to element 1 (here hydrogen).

$V'_{1i}$ is given by :

$$
V'_{1i} = \left( 4 \pi \rho r^2 \right) V_{1i}
$$

with :

$$
V_{1i} = -D_{1i} \left[ \left( A_i - \frac{Z_i}{2} - \frac{1}{2}
  \right) \left( \frac{m_H}{kT} \frac{GM}{R^2} \right) - \alpha _{1i}
  \nabla \ln T \right]
$$

The thermal diffusion coefficient $\alpha _{1i}$ is computed using the
formalism of Paquette et al. 1986; $A_i$ and $Z_i$ represent the atomic
mass number and charge of element i and $m_H$ is the atomic hydrogen
mass.$M$ and $R$ stand for the stellar mass and radius.

A normalisation condition on the mass fractions of all the elements has
to be added
to correct for the
center-of-mass displacement.

The diffusion equation has to be solved simultaneously for
all the considered elements. The order of magnitude of the time scales
generally implies the computation of many iterations of the diffusion
process for a single evolutionary time step. For each computation of a
new model along the evolutionary track, the tables of abundances inside
the star have to be transferred for every element, as a function of the
internal mass. For the model consistency, these abundance profiles must
be taken into account in the interpolation of the opacity tables.

For the Sun, the whole process of complete time evolution
has to be iterated several times from the
beginning, with small adjustments in the original helium mass fraction and
mixing length parameter, to obtain the right Sun and the right age
(luminosity and radius with a relative 
precision of at least $10^{-4}$).

\subsection{The Treatment of Collisions}

The diffusion time scales are direct functions of the collision
probabilities for the considered species. A good treatment of collisions
is thus necessary to obtain the abundance variations with a high degree
of precision.

For the diffusion of neutral atoms in a neutral gas, the ``hard sphere
approximation" is used. For ions moving in a neutral medium, or neutrals
moving
in a plasma,  the polarisation of the neutrals have to be taken into
account.
For collisions between charged ions, problems similar to those
encountered for the equations of state have to be solved. The basic
question concerns the divergence of the coulomb interaction cross
sections. In the first computations of diffusion, the ``Chapman and
Cowling approximation" was used, assuming a cut-off of the cross section
equal to the Debye shielding length. Average values of the shielding
factor were used for analytical fits of the resulting diffusion
coefficients.

Paquette et al. (1986) proposed a more precise treatment of this
problem. 
They pointed out that
the Debye shielding length has no physical meaning as soon as it is
smaller than the inter-ionic distance. 
They proposed to introduce a screened coulomb
potential with a characteristic length equal to the largest of the Debye
length and inter-ionic distance, and they gave tables of collision
integrals which can be used in the computations of diffusion processes
in the stellar gases. The Paquette et al. approximation should be
generally used
in the computations of stellar structure. It may however in most cases
be replaced by an analytical expression given by Michaud and Proffitt
(1992).

\subsection{the radiative acceleration}

Many authors have computed the gravitational and thermal diffusion of
helium and heavier elements in the Sun with various approximations :
see, for example,
Michaud and Vauclair (1991) and references therein; Cox, Guzik and 
Kidman (1989); Bahcall and Pinsonneault (1992); Proffitt (1994); 
Thoul, Bahcall and Loeb (1994); Christensen-Dalsgaard, Proffitt and
Thompson (1993); RVCD. In all cases the radiative accelerations were
neglected.

For the first time, Turcotte et al (1998) have consistently computed the
radiative accelerations on the elements included in the OPAL opacities.
They have found that, contrary to current belief, the effect of
radiation can, in some cases, be as large as $\cong~40\%$ that of
gravity below the solar convective zone. This is important only for
metals however, and not for helium. When the radiative accelerations are
neglected, the abundances of most metals change by $\cong~7.5\%$
if complete ionisation is assumed below the convection zone, and by
$\cong~8.5\%$ if detailed ionisation rates are computed. When the
radiative accelerations are introduced, with detailed ionisation, the
results lie in-between. The resulting effect on the solar models is
small and can be neglected (while it becomes important for hotter
stars).

\section{Evidence of Element Settling inside the Sun from
Helioseismology}

Solar models computed in the old ``standard" way, in which the element
settling is totally neglected, do not agree with the inversion of the
seismic modes. This result has been obtained by many authors, in
different ways (see Gough et al (1996) and references therein). There is
a characteristic discrepancy of order one percent, just below the convective
zone, between the sound velocity computed in the models and that of the
seismic Sun. Introducing the element settling reconciliates the two
results. 
Fig.1 shows an example of solar models obtained with and without element
settling in the computation.
These models have been obtained with the Toulouse code, as described in
Charbonnel, Vauclair and Zahn (1992) and RVCD. 
Some improvements have been introduced in the treatment of the opacities
and equation of state, as described in Richard, Vauclair and Charbonnel
(1998).

These computations are compared to the results of helioseismology in
collaboration with the Warsaw group (Dziembowski et al., 1994). 
The values of the function $u = P/\rho$ as obtained from the
inversion
of the solar oscillation modes (seismic Sun) are displayed
together with the results of our models.

\begin{figure}[h]
\epsfysize=8cm
\epsfbox{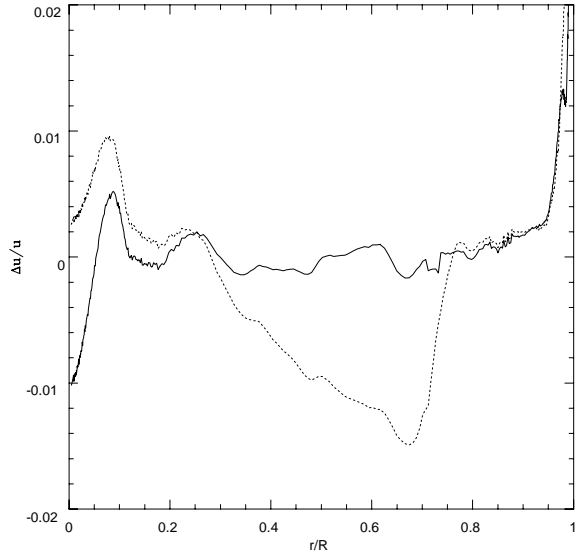} 
\caption{Comparison of the $u = P/\rho$ function in the ``seismic Sun" 
and in our models.
Dotted line: without element settling; solid line: with settling 
computed for helium and 14 other elements (from Richard,
 Vauclair and Charbonnel, 1998)}
\end{figure}

It could be possible to reduce the discrepancy between the sound
velocity in the old solar standard models and in the seismic Sun by
adding other effects than element settling. For example changes in the
opacities could possibly lead to similar results. However, as already
pointed out, element settling must not be considered as a new parameter
added in the computations. It represents second order effects in the
physics of auto-gravitational spheres, precisely and without any free
parameter. Introducing element settling in the standard models means
improving the physics. The fact that these new models lie closer to the
seismic Sun than the old ones is quite encouraging and may be considered
as a proof that the physical improvements are correct.

\section{Discussion : necessity of mild mixing, $^7$Li and $^3$He}

Although the introduction of pure element settling in the solar models
considerably improves the consistency with the ``seismic Sun", some
discrepancies do remain, particularly below the convective zone where a
"spike" appears in the sound velocity (see Gough et al. 1996).
The helium profiles
directly obtained from helioseismology (Basu 1997, Antia and Chitre
1997) show indeed a helium gradient below the convection zone which is
smoother than the gradient obtained with pure settling. Furthermore,
standard solar models including element settling do not reproduce the
observed abundances of lithium.

The abundance determinations in the solar photosphere show that lithium
has been depleted by a factor of about 140 compared to the protosolar
value while beryllium
is generally believed to be depleted by a factor 2.
These values have widely
been used to constraint the solar models (e.g. RVCD).
However, while the lithium depletion factor seems well established, the
beryllium value is still subject to caution. Balachandran and Bell
(1998) argue that the beryllium depletion is not real due to an
underestimate of the 
opacity of the continuum in the abundance
determinations. Their new treatment leads to a solar value identical to the
meteoritic value.

In RVCD, a mild mixing below the convection zone, attributed to
rotation-induced shears (Zahn 1992), was introduced to account for the
lithium and beryllium depletion. It was shown that such a mixing may
also wipe out the spike in the sound velocity, leading to more
consistent solar models than the standard ones, computed with pure
element settling. In these models the abundance of $^3$He
increased in the convection zone, due to a small dredge up from the tail
of the $^3$He peak.

Observations of the $^3$He/$^4$He ratio in the solar
wind
and in the lunar rocks (Geiss 1993,
Gloecker and Geiss 1996, Geiss and Gloecker 1998) show that this ratio
may not
have increased by more than $\cong~10\%$ since 3 Gyr in the Sun, which
is in contradiction with the results of RVCD.
While the occurence of some mild mixing below
the solar convective zone is needed to explain
the lithium depletion and helps for the conciliation of the models with
helioseismological constraints,
the $^3$He/$^4$He
observations put a strict constraint on its efficiency.

Vauclair and Richard (1998) have tried several parametrizations of
mixing below the solar convection zone, which could reproduce both the
$^7$Li and the $^3$He constraints. The only way to obtain such a result
is to postulate a mild mixing, which would be efficient down to the
lithium nuclear burning region but not too far below, to preserve the
original $^3$He abundance. 
A mixing effect decreasing with time, as
obtained with the rotation-induced shear hypothesis, 
helps to obtain a $^7$Li destruction without increasing $^3$He
too much
, as the $^3$He peak itself builts up during the solar life.
A ``cut-off" od the mixing process must however be postulated at some
depth just below the $^7$Li destruction layer.

The various kinds of mixing processes which may take place below the
solar convection zone are summarized in J.P. Zahn's review (this
conference). Here we have tested an effective diffusion coefficient
varied as a parameter that we have adjusted to reproduce the observed
abundances. We found that the observations of $^7$Li and $^3$He are well
accounted for with a very mild mixing 
described by a diffusion coefficient not larger than 
$10^{3}$.cm$^2$.s$^{-1}$,
vanishing at about two scale heights below the bottom of the convection
zone (figure 2).
In this case $^7$Li is destroyed by a factor 140 and $^3$He is not
increased by more than $\cong~5\%$.
Meanwhile, beryllium is not
destroyed by more than $\cong~20\%$.

\begin{figure}[h]
\epsfysize=8cm
\epsfbox{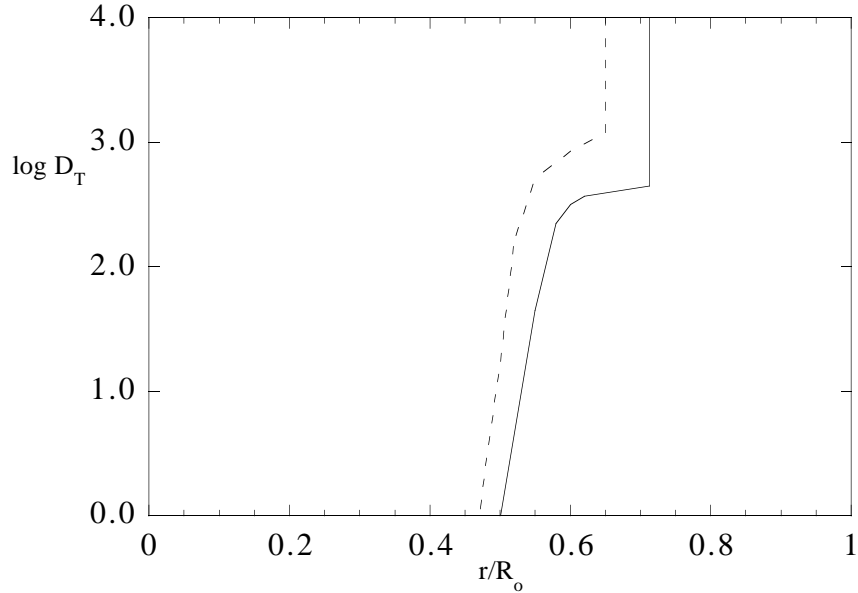}
\caption{Example of profiles of the macroscopic diffusion coefficient below the
solar convection zone, which can account for both the $^7$Li and $^3$He
constraints (after Vauclair and Richard, 1998). 
In this case D$_T$ decreases with time. The profiles are shown at 1.22 Gyr
(dashed line) and 4.6 Gyr (solid line).
$^7$Li is then
depleted by a factor 140 while $^3$He does not increase by more than
$5\%$ during the whole solar life. 
In these models, beryllium is not depleted 
by more than $20\%$}
\end{figure}

In summary the best solar models must include the effect of element
settling, which represents an improvement on the physics, without any
free parameter added. These models can be considered as the new
``standard" models. They cannot however reproduce the $^7$Li depletion
and they lead to a spike in the sound velocity, compared to the seismic
Sun, just below the convective zone. These two observations suggest the
presence of some mild mixing in this region of the internal Sun. Adding
the very strict constraint on the abundance of $^3$He as given by Geiss
and Gloecker (1998) leads to a precise description of the allowed
profile of the macroscopic diffusion coefficient below the convective
zone. This result can be taken as a challenge for the hydrodynamicists.

\bigskip

{\bf References}

\begin{itemize}
\item  {}Antia, H.M., Chitre, S.M..: 1997, {\it A\&A}, {\bf 000},000
\item  {}Bahcall, J.N., Loeb, A.: 1990, {\it ApJ} {\bf 360}, 267
\item  {}Bahcall, J.N., Pinsonneault, M.H.: 1992, {\it Reviews of
Modern Physics} {\bf 64}, 885
\item  {}Balachandran, S., Bell, R.A. : 1998, preprint
\item  {}Basu, S.: 1997, {\it Mon. Not. R. Astron. Soc.} {\bf 288},
572
\item  {}Burgers, J.M..: 1969, {\it Flow Equations for Composite
Gases}, 
New York : Academic Press
\item  {}Charbonnel, C.,  Vauclair, S., Zahn, J.P.: 1992, {\it A\&A}
{\bf 255}, 191
\item  {}Chapman, S., Cowling,T.G.: 1970, {\it The mathematical Theory
of Non-Uniform Gases}, Cambridge University Press, 3rd ed.
\item  {}Christensen-Dalsgaard, J., Proffitt, C.R., Thompson, M.J.:
1993, {\it ApJ} {\bf 408}, L75
\item  {}Cox, A.N., Guzik, J.A., Kidman, R.B.: 1989, {\it ApJ}
{\bf 342}, 1187
\item  {}Dziembowski, W.A., Goode, P.R., Pamyatnikh, A.A.,
Sienkiewicz, R.: 1994, {\it ApJ} {\bf 432}, 417
\item  {}Geiss, J.: 1993, {\it Origin and Evolution of the Elements},
ed. Prantzos, Vangioni-Flam \& Cass\'e (Cambridge Univ.
Press), {\bf 90}
\item  {}Gloecker, G., Geiss, J.  : 1996, {\it Nature}, {\bf 381}, 210
\item  {}Geiss, J., Gloecker, G. : 1998, {\it Space Sci. Rev.}, this
volume
\item  {}Gough,D.O., Kosovichev, A.G., Toomre, J., Anderson, E., Antia,
H.M., Basu, S.,  Chaboyer, B., Chitre, S.M., Christensen-Dalsgaard, J.,
Dziembowski, W.A., Eff-Darwich, A., Elliott, J.R., Giles, P.M., Goode,
P.R., Guzik, J.A., Harvey, J.W., Hill, F., Leibacher, J.W., Monteiro,
M.J.P.F.G., Richard, O., Sekii, T., Shibahashi; H., Takata, M.,
Thompson, M.J., Vauclair, S., Vorontosov, S.V. : 1996, {\it Science}
{\bf 272}, 1296
\item  {}Michaud, G.: 1970, {\it ApJ} {\bf 160}, 641
\item  {}Michaud, G., Proffitt,C.R.  1992, in {\it Inside the Stars},
ed. W.W.Weiss and A. Baglin, vol. 371, IAU, (San Francisco: PASPC), pp
246-259
\item  {}Michaud, G., Vauclair, S.: 1991, in {\it Solar Interior and
Atmosphere}
A.N. Cox, W.C. Livingston, M.S. Matthews ed., The University of Arizona
Press, 
p. 304
\item  {}Paquette, C., Pelletier, C., Fontaine, G., Michaud, G.: 1986,
{\it ApJS} {\bf 61}, 177
\item  {}Proffitt, C.R.: 1994, {\it ApJ} {\bf 425}, 849
\item  {}Proffitt, C.R., Michaud, G. : 1991, {\it ApJ} {\bf 380}, 238
\item  {}Proffitt, C.R., VandenBerg, D.A. : 1991, {\it ApJS} {\bf 77},
473
\item  {}Richard, O., Vauclair, S., Charbonnel, C. : 1998, preprint
\item  {}Richard, O., Vauclair, S., Charbonnel, C., Dziembowski,
W.A.: 1996, {\it A\&A} {\bf 312}, 1000
\item  {}Richer, J., Michaud,G. : 1993, {\it ApJ} {\bf 416}, 312
\item  {}Thoul, A.A., Bahcall, J.N., Loeb, A.: 1994, {\it ApJ} {\bf
421}, 828
\item  {}Turcotte,S., Richer,J., Michaud,G., Iglesias,C.A., Rogers,F.J.
: 1998, {\it ApJ} {\bf000}, 000
\item  {}Vauclair, S. : 1998, {\it Space Sci. Rev.}, in press
\item  {}Vauclair, S., Vauclair, G.: 1982, {\it ARA\&A} {\bf 20}, 37
\item  {}Vauclair, S., Richard, O.: 1998, preprint
\item  {}Zahn, J.P.: 1992, {\it A\&A} {\bf 265}, 115

\end{itemize}

\end{document}